\documentclass[runningheads]{llncs}
\usepackage{bm}
\usepackage{amsfonts,amssymb,amsmath}
\usepackage{graphicx}
\usepackage{xcolor}
\usepackage{url}
\usepackage{epstopdf}
\usepackage{array}
\usepackage{booktabs}
\usepackage{multirow}
\usepackage{subfigure}

\begin{document}

\title{3D U-Net Based Brain Tumor Segmentation and Survival Days Prediction\thanks{This work is supported by National Natural Science Foundation
(NNSF) of China under Grant 61871420. \protect\\The source code of this work is opened on https://github.com/woodywff/brats\_2019}}

\author{Feifan Wang\inst{1}\orcidID{0000-0002-9525-0656} \and
Runzhou Jiang\inst{1}\orcidID{0000-0002-9505-1112}\and\\
Liqin Zheng\inst{1}\orcidID{0000-0003-2486-4295}\and
Chun Meng\inst{1}\orcidID{0000-0003-0145-6627}\and
Bharat Biswal\inst{1,2}\orcidID{0000-0002-3710-3500}}

\authorrunning{F. Wang et al.}


\institute{Center for Information in Medicine, School of Life
Science and Technology, University of Electronic Science
and Technology of China, Chengdu 611731, China
\email{woodywff@aliyun.com, woodywff@uestc.edu.cn}
\and
Department of Biomedical Engineering, New Jersey Institute of Technology, Newark, NJ, USA\\
\email{bbiswal@gmail.com}}
\maketitle              

\begin{abstract}
Past few years have witnessed the prevalence of deep learning in many application scenarios, among which is medical image processing. Diagnosis and treatment of brain tumors requires an accurate and reliable segmentation of brain tumors as a prerequisite. However, such work conventionally requires brain surgeons significant amount of time. Computer vision techniques could provide surgeons a relief from the tedious marking procedure. In this paper, a 3D U-net based deep learning model has been trained with the help of brain-wise normalization and patching strategies for the brain tumor segmentation task in the BraTS 2019 competition. Dice coefficients for enhancing tumor, tumor core, and the whole tumor are 0.737, 0.807 and 0.894 respectively on the validation dataset. These three values on the test dataset are 0.778, 0.798 and 0.852. Furthermore, numerical features including ratio of tumor size to brain size and the area of tumor surface as well as age of subjects are extracted from predicted tumor labels and have been used for the overall survival days prediction task. The accuracy could be 0.448 on the validation dataset, and 0.551 on the final test dataset.

\end{abstract}

\section{Introduction}
Human brain stays in a delicate balance under the enclosure of the skull. A brain tumor is a bunch of abnormal brain cells that may harass the balance~\cite{brain_0}. Primary brain tumors originate in the brain, while others belong to the secondary or metastatic brain tumors that come from other organs. Brain tumors can also be categorized as malignant or benign, the former are cancerous and easy to be spread to the other part of the brain while the later not. Nevertheless, in both cases, the growth of brain tumor in rigid brain space could result in a dysfunction or even life-threatening symptom for human body. Depending on the size and location of the tumor, people may have different symptoms caused by the growing of tumor cells. Some tumors would invade brain tissue directly and some cause pressure on the surrounding brain. As a result, people may suffer from vomiting, blurred vision, confusion, seizures, et al. Magnetic Resonance Imaging (MRI) and resection surgery are the most common diagnosis and treatment means respectively currently used for brain tumors~\cite{brain_1}. A priority for a neurosurgeon is to mark the tumor region precisely. Too much or too less surgery may give rise to more loss and suffering. Unfortunately, manually labeling is a laborious and time consuming work for a doctor. Moreover, because of inevitable practical operation factors, it is difficult to replicate a segmentation result exactly the same.

Determining the best computer assistant solutions to the segmentation task, Multimodal Brain Tumor Segmentation Challenge 2019 provides ample MRI scans of patients with gliomas, the most common primary brain tumor, before any kind of resection surgery~\cite{brats_1,brats_2,brats_3}. For training datasets, 259 subjects with high-grade gliomas (HGG) and 76 subjects with low-grade gliomas (LGG) were used~\cite{brats_4,brats_5}. Each subject had four $240\times 240\times 155$ structural MRI images, including native (T1), post-contrast T1-weighted (T1Gd), T2-weighted (T2), and T2 Fluid Attenuated Inversion Recovery (FLAIR) volumes. Meanwhile, pathologically confirmed segmentation labels for each subject also come as $240\times 240\times 155$ images with values of 1 for the necrotic (NCR) and the non-enhancing tumor core(NET), 2 for edema (ED), 4 for enhancing tumor (ET), and 0 for everything else. Further, the segmentation task defines three sub-regions for evaluation, they are 1) the tumor core (TC) including NCR, NET and ET, 2) the ET area, 3) the whole tumor (WT) which is the combination of TC and ED. All these provided MRI images were collected from 19 institutions and had undergone alignment, $1\times1\times1$ mm resolution resampling and skull stripping. Another task in BraTS 2019 is to predict the overall survival (OS) days of patients after the gross total resection (GTR) surgery. All the OS values are provided together with the age of patients with resection status of GTR.

Deep learning has reentered prosperous ever since AlexNet won the ImageNet competition in 2012~\cite{alexnet,imagenet}, which to a great extent attributes to the massively ascending dataset scale and computing power. The advancement of convolutional neural networks came up with a lot of crafted deep learning designs, like VGG-Net~\cite{vgg}, Inception networks~\cite{inception,inception-v4} and ResNet~\cite{resnet}. These crafted architectures together with advanced open source frameworks like tensorflow and pytorch energize the development in many research and industrial fields. Semantic segmentation in image processing is to separate the target object from other areas. Fully convalutional networks (FCN) empowers CNN to be able to label each pixel by means of a plain upsampling idea~\cite{fcn}. For medical images, usually they don't share the same features with ordinary pictures from dataset like ImageNet or CIFAR-10/100~\cite{imagenet,cifar10}, which makes it difficult for pre-trained networks on those datasets to be directly used and leaves spaces for specific inventions. U-Net stood out from the IEEE International Symposium on Biomedical Imaging (ISBI) challenge that segments electron microscopy images of the drosophila cell~\cite{unet}. When it comes to 3D volumetric medical images, the inventors of U-Net also proposed feasible solutions~\cite{3d-unet}. 

BraTS initiated by Center for Biomedical Image Computing and Analytics (CBICA) encourages participants to identify competitive solutions to brain tumor segmentation tasks. Most former teams took use of U-Net or put it in ensemble with other modules. For example, the 1st ranked model of BraTS 2018 added a variational auto-encoder (VAE) branch on the U-Net structure to give a new regularization item in the loss function~\cite{nvidia-net}. Isensee et al. argued that a well trained U-Net could be powerful enough and brought out a fine-tuned U-Net model that won the 2nd place in the contest~\cite{isensee-net}. Moreover, the second task of BraTS is to predict how many days could a patient survive after the GTR operation. Previously, the best record was obtained by performing a linear regression model developed by Feng et al. in BraTS 2018~\cite{feng_os}, the features they used include the size and length of tumors and the age information.

In this work, we illustrate our solutions to the two tasks in BraTS 2019. Four modalities(T1, T1Gd, T2, and T2-FLAIR) of structural MRI images collected from patients with gliomas are processed and fed into the network. 
\begin{figure}[!htbp]
  \centering
\includegraphics[width=0.55\textwidth]{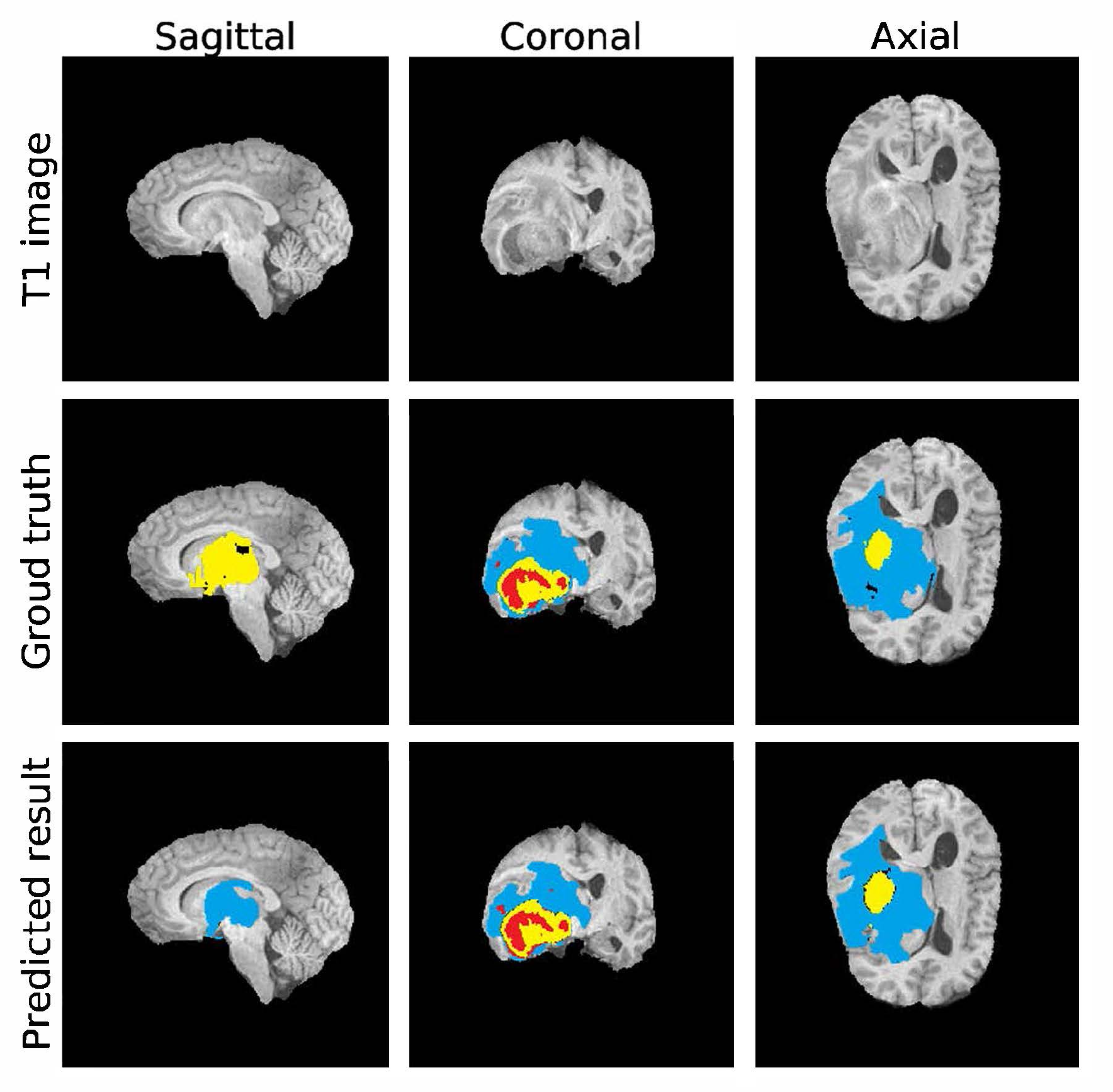}
  \caption{An example of one modality of input data and the corresponding ground truth and predicted labels, each picture illustrates a slice of 3D MRI images. For ground truth and predicted result, four colors represent four label values. Red for value 1 (NCR and NET), blue for value 2 (ED), yellow for value 4 (ET), black for value 0 (everything else). (Notice that brain areas other than tumor part are also supposed to be black in the six label figures, we draw the brain image just for illustration purpose.)}
  \label{fig0}
\end{figure}
In particular, rather than normalizing voxel values on the whole image with black background, only the brain wise area has been taken into account for the normalization and scaling. Two phases of training with different patching strategies were undertaken, with one keeping an eye on the black background and the other not, both with a patch covering the center field. An extra parameter was designed for each image to remember the minimum cube that could encapsulate it. All the patching and recovery procedures were maneuvered based on such cubes.
Different kinds of tumor tissues would be labeled with different values, as can be seen in Fig.~\ref{fig0} which gives an example of three slices of one T1 brain image in directions of segittal, coronal and axial. These segmentation results have been analyzed and further used for the overall survival task. The proportion of tumor compared to the whole brain and the ratio of length of one sub-region to another have been extracted, together with the phenotypic age information, as the input features.

\section{Method}

\subsection{Preprocessing}

All the structural MRI images have been bias field corrected through N4ITK Insight Toolkit hopefully to minimize the bias caused by the differences in multiple scanners, institutions and protocols~\cite{n4itk}. Normalization was performed for each modality, by accumulating the voxel values inside brain skull throughout all the training images and calculating the mean value $\mu$ and the standard deviation $\sigma$. Given $\bm{A}\in \mathbb{R}^{240\times240\times155}$ represents an original image, the z-score normalization and min-max scaling are deployed in the following manner. 

\begin{align}
	\hat{A}_{ijk} &=
    \begin{cases}
        (A_{ijk}-\mu)/\sigma & \text{if $A_{ijk}\not=0$} \\
        \qquad\ 0 & \text{else,}
    \end{cases}\label{eq1}\\
    \tilde{A}_{ijk}&=
    \begin{cases}
    	100\times\bigg(\dfrac{\hat{A}_{ijk}-\hat{A}_{\rm{min}}}
        {\hat{A}_{\rm{max}}-\hat{A}_{\rm{min}}}
                         +0.1\bigg) & \text{if $A_{ijk}\not=0$}\\                         
        \qquad\   0  & \text{else,}  
    \end{cases}\label{eq2}
\end{align}
in which $\hat{A}_{\rm{min}}$ and $\hat{A}_{\rm{max}}$ indicate the minimum and maximum values for all the $\hat{A}_{ijk}$ with corresponding $A_{ijk}\not=0$, $i,j\in [0,239]$, $k\in [0,154]$.
Brain area voxel values in the finally preprocessed image would range from 10 to 110, which are discriminated from background value 0. Notice that for validation and test datasets, we still use the $\mu$ and $\sigma$ calculated from the training dataset for the z-score normalization. 

\subsection{Patching strategies}
\begin{figure}[!htbp]
  \centering
  \subfigure[]{
    \label{fig1a}
    \includegraphics[width=0.45\textwidth]{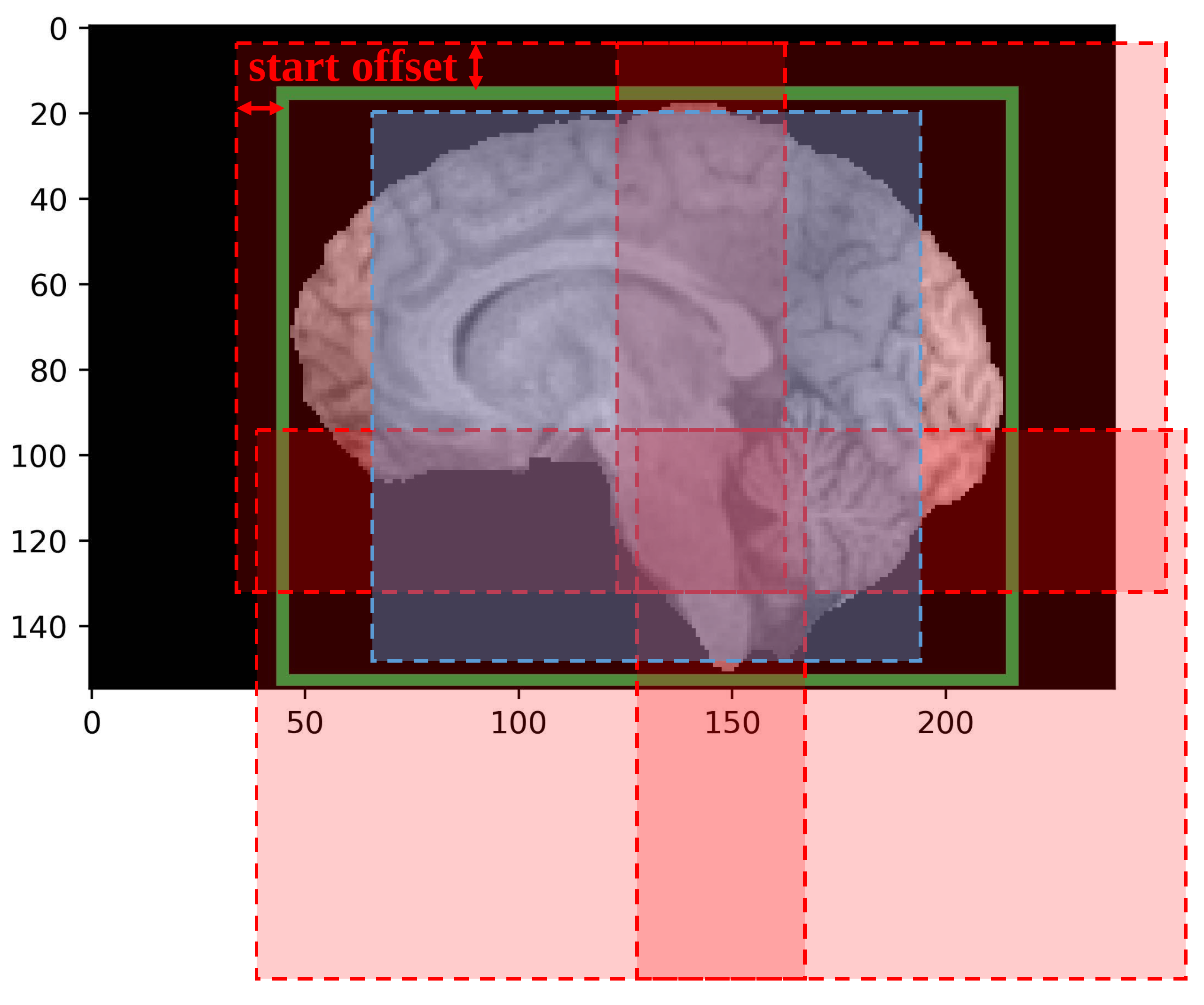}}
  \hspace{0.0in}
  \subfigure[]{
    \label{fig1b}
    \includegraphics[width=0.45\textwidth]{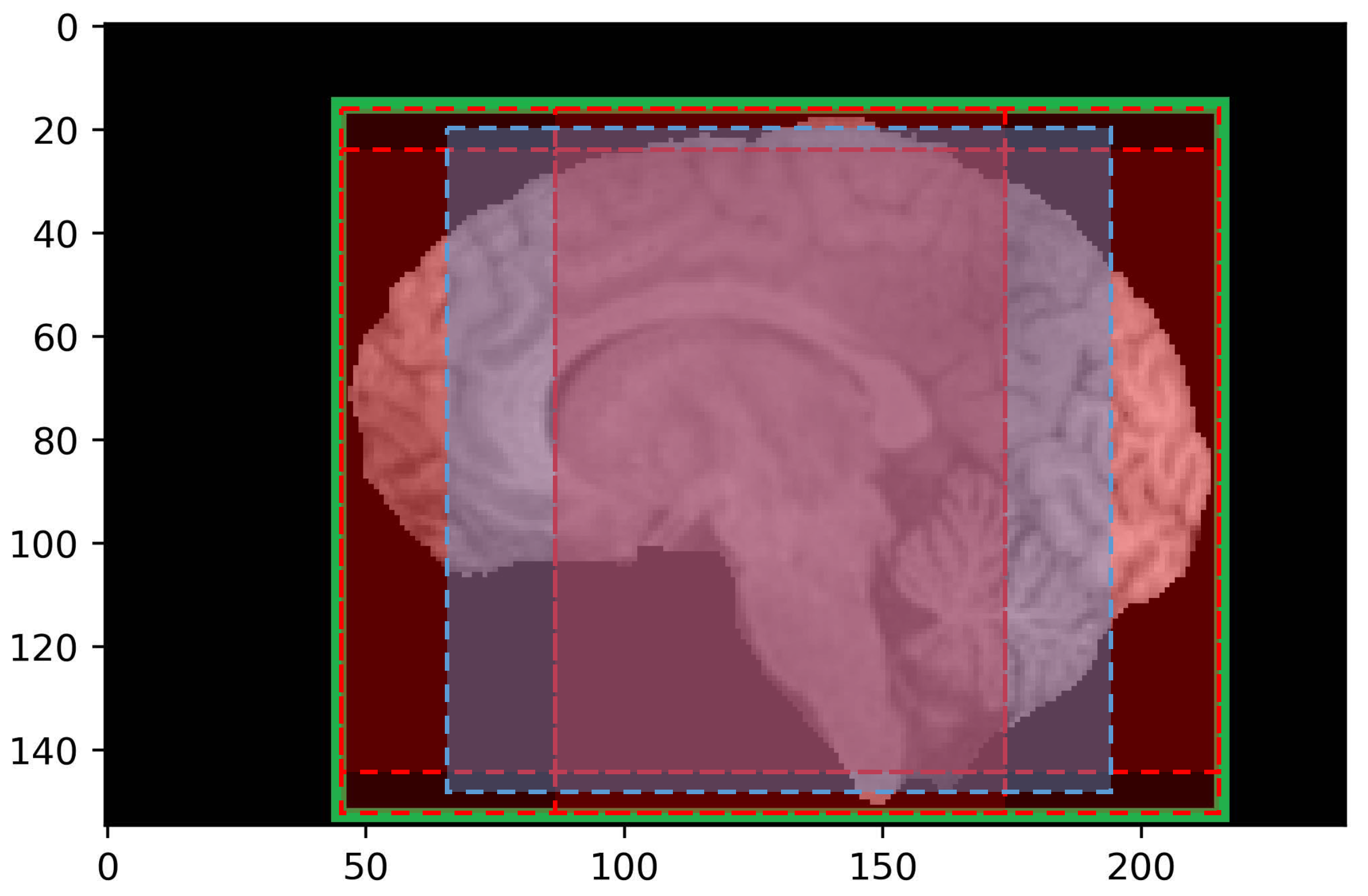}}
  \caption{Patching strategies. The green rectangle represents the boundary of the brain, red and blue dashed cubes indicate the patches, in particular, blue for the fixed center one. Start offset marks the largest distance to the boundary a patch could start with. (\textbf{a}) Patching strategy in the first training phase. (\textbf{b}) Patching strategy in the second training phase.}
  \label{fig1}
\end{figure}

Patching strategy would make it possible for less powerful GPU to deal with a large image. 
In this work we leverage two kinds of patching strategies, both were maneuvered based on the cuboid boundary of the brain, as shown in Fig.~\ref{fig1}, patch size is $128\times128\times128$. For the convenient of drawing, we use a 2D picture to illustrate the idea behind operations on 3D images.

For the first strategy, as seen in Fig.~\ref{fig1a}, a cubic patch starts with a random distance between 0 to 4 voxels away from the border. The overlap of each two neighbor patches is 32, which stands for each time when the patching window moves 96 voxels. This strategy would generate patches with a bunch of background values. For the second image displayed in Fig.~\ref{fig1b}, all the patches were arranged to the largest extent inside the brain area, each corner of the quadrilateral boundary has a corresponding patch with one corner completely matched. For the training process, two of the patching strategies are employed in sequence. The second strategy is also the one used for the prediction procedure. One extra blue dashed cube that exists in both figures in Fig.~\ref{fig1} refers to a patch we fix in center of the brain area for each image, considering the large amount of information in this part.

\subsection{Models}
\subsubsection{Segmentation task:}
\begin{figure}[!htbp]
\centering
\includegraphics[width=0.8\textwidth]{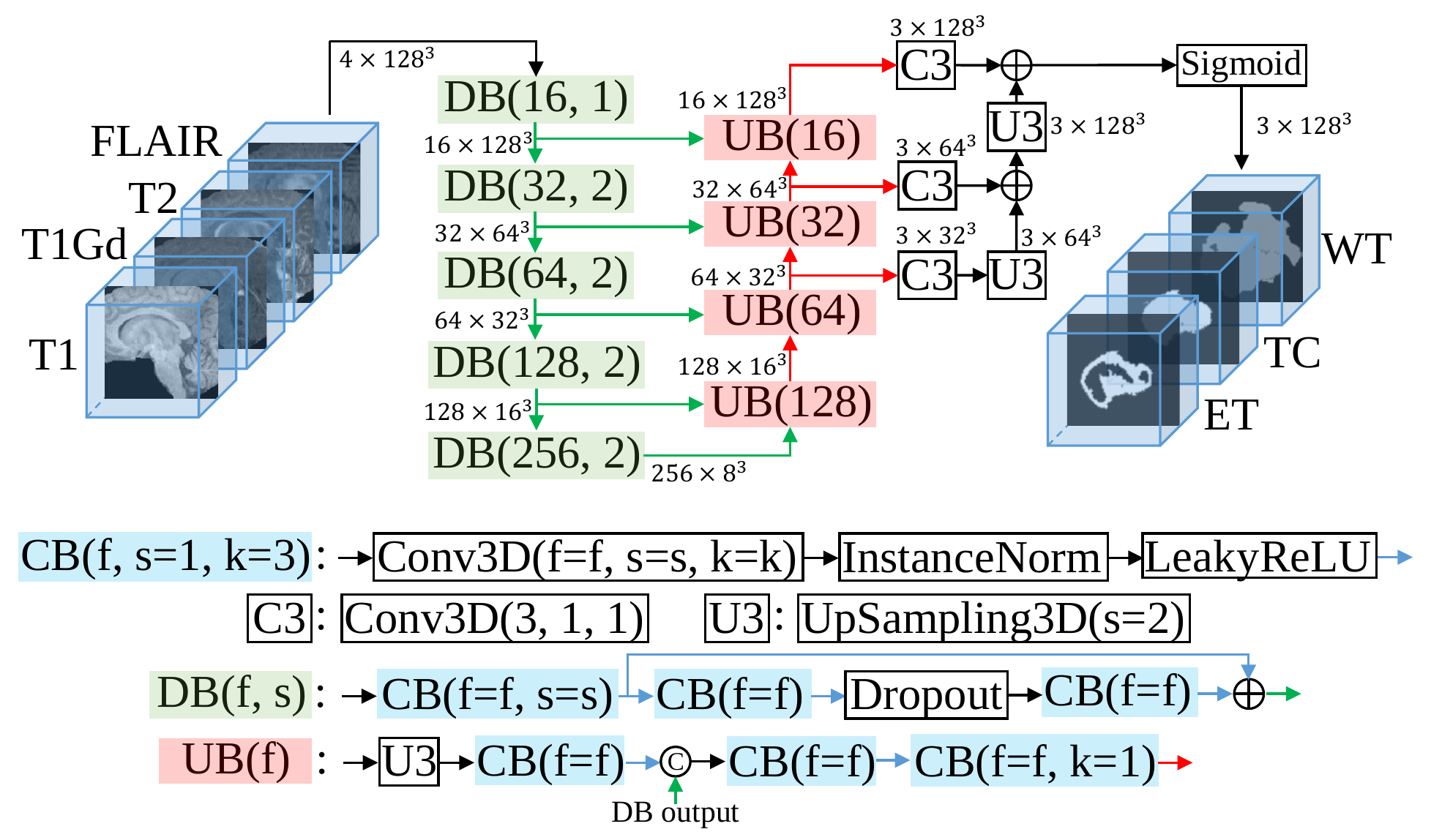}
\caption{Schematic of the network for segmentation task. Self-defined modules are listed downstaires. CB refers to Convalutional Block, which takes 3 parameters from Conv3D block inside: $f$ is the number of filters, $s$ is the step length, and $k$ is the kernel size. C3 is short for Conv3D module and U3 for UpSampling3D with upsampling factor equals 2. DB and UB mean Downward Block and Upward Block respectively. In general, black rimmed blocks are original modules in frameworks, while colored blocks and arrows indicate the developed ones. $\oplus$ is element-wise addition, $\copyright$ means concatenation. } \label{fig2}
\end{figure}

Following the framework of U-Net by Isensee et al.~\cite{isensee-net}, the architecture of the segmentation network has been depicted in Fig.~\ref{fig2}. 
It takes $4\times128\times128\times128$ matrices as input, each of which is stacked by 4 $128\times128\times128$ patches of different modalities. The Downward Block (DB) squeezes the image size and stretches the channel length. Particularly, as the basic feature embedding element, the Convalutional Block (CB) works with the instance normalization and leaky ReLU modules~\cite{instance_norm}. As in ResNet, DB also bypasses the front message to the end to fight with the weight decay and overfitting problems. Upward Blocks (UB) are in charge of reconstructing the location information by means of concatenating the corresponding DB output. They recover the image size to $128\times128\times128$ and shrink the depth of channel to 3, each of which is the probability matrix that demonstrates the confidence of each voxel belonging to one certain sub-region of the tumor.

The weighted multi-class Dice loss function has been proved to be efficient in former BraTS competitions~\cite{isensee_2017}. As exhibited in Eq.~(\ref{eq3}),

\begin{equation}
    L=-\sum_{c=0}^{2}\frac{\sum_{i,j,k=0}^{127}Y_{cijk}\hat{Y}_{cijk}}
            {\sum_{i,j,k=0}^{127}Y_{cijk}+\sum_{i,j,k=0}^{127}\hat{Y}_{cijk}},
\label{eq3}
\end{equation} 
in which $\bm{Y}$ indicates the $3\times128\times128\times128$ matrix generated from the ground truth image, $\hat{\bm{Y}}$ represents the output from the constructed network.

The 3 channels in the output would be mixed to one $128\times128\times128$ image that each voxel chooses to be the value of one tumor sub-region label or to be zero as the background (by setting up one threshold). Meanwhile, the priority of ET is higher than TC's which in turn is higher than that of WT, which means we compare priorities rather than probabilities once the probability value is over threshold.  
At the end of the prediction, all the labeled patches with numbers indicating the tumor sub-regions would be concatenated into a brain boundary (the green rectangle in Fig.~\ref{fig1a}) sized image, in which the overlapped part with more than one values would take the average of them. This image would be further recovered into the original size of $240\times240\times155$  according to the saved boundary information.

\subsubsection{Overall survival days prediction:}
The predicted tumor images would be further utilized for the estimation of the OS days. In this task, we select seven features from numerous candidates. The first three are the ratios of the volume of each tumor sub-region to the size of the whole brain. Then we calculate the gradient of each kind of tumor matrix and sum up non-zero gradient values to approximate the area of the tumor surface. Last but not the least, age information of subjects have been taken into account. The combination of these seven features have been demonstrated with better performance than other mixtures. 
The model we choose to solve this problem is straight forward---a fully connected neural network with two hidden layers, each with 64 filters.

\section{Results}
All the programs developed for this work were written in python and Keras backend by tensorflow, on a single GTX 1080ti GPU installed desktop.

\begin{figure}[!htbp]
\centering
\includegraphics[width=0.9\textwidth]{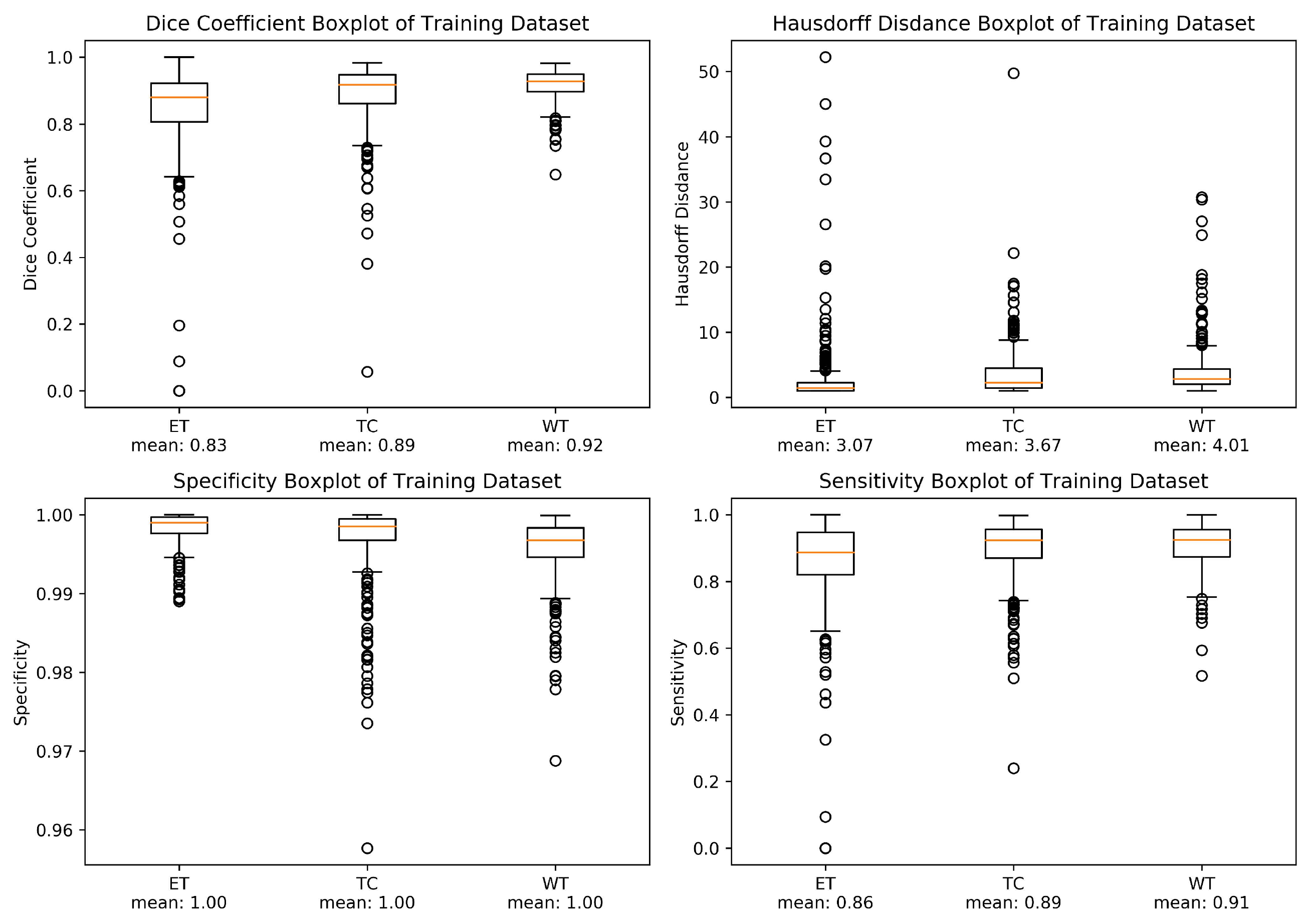}
\caption{Dice coefficient, specificity, sensitivity and hausdorff for training dataset.} \label{fig3}
\end{figure}

\begin{figure}[!htbp]
\centering
\includegraphics[width=0.9\textwidth]{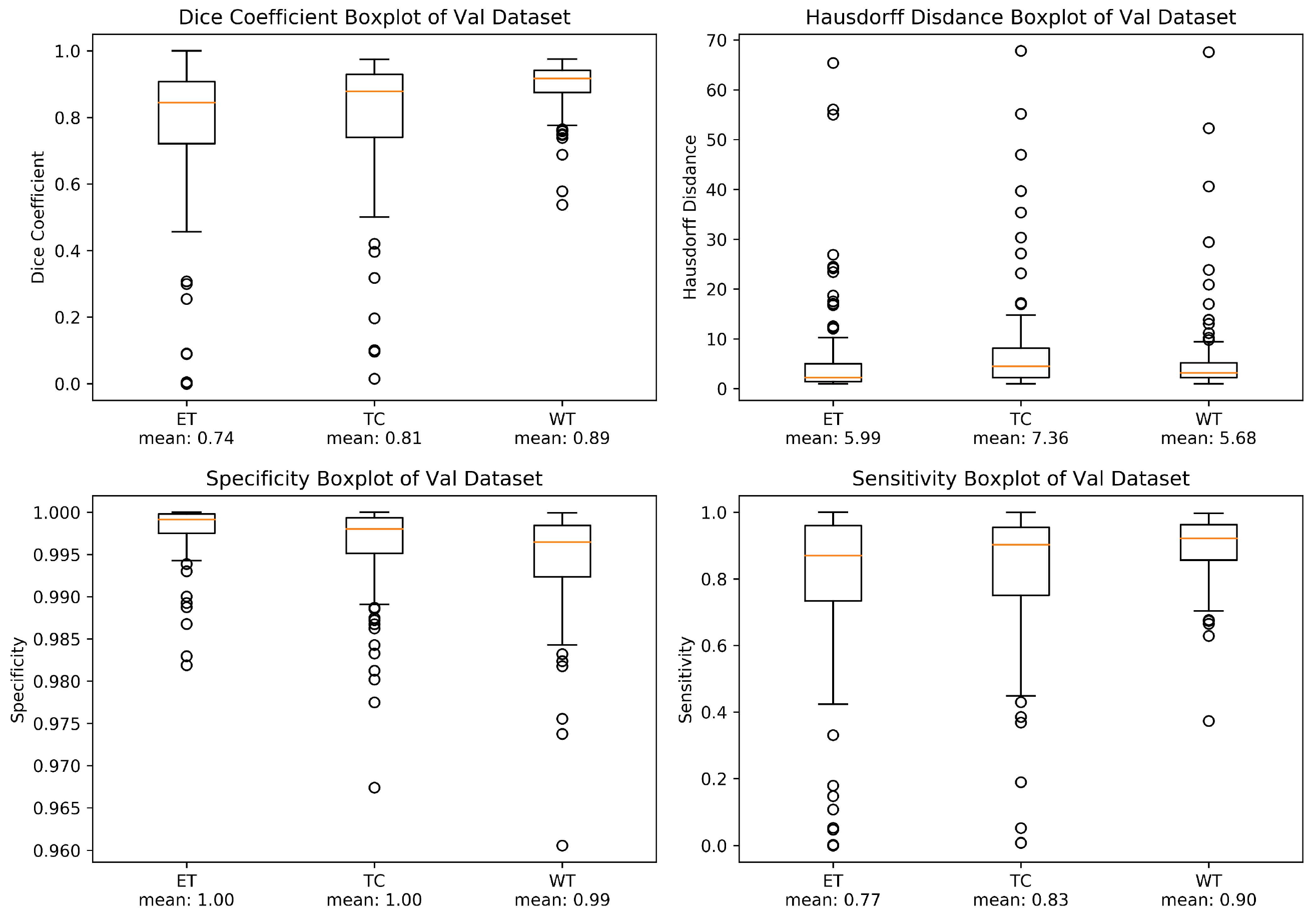}
\caption{Dice coefficient, specificity, sensitivity and hausdorff for validition dataset.} \label{fig4}
\end{figure}

In segmentation task, there are 335 labeled subjects for training, 125 unlabeled subjects for validation and 166 unlabeled subjects for the final test. We trained the network with two patching strategies each for 100 epochs, the second phase use the first phase saved model as pre-trained. Start offset is 4, patch overlap is 32, initial learning rate of Adam optimizer is 5e-4, the learning rate would drop by 50\% after 10 steps without loss value decrease. Data augmentation has been undertaken on the fly for each patch, including random rotation, flip and distortion. Table~\ref{tab1} exhibits the mean values of all the required criteria in BraTS 2019, Fig.~\ref{fig3} and Fig.~\ref{fig4} illustrate more details of that. Because we are only afforded the summary stats of results on the final test dataset, here we just illustrate the mean values of the dice coefficient and Hausdorff distance in Table~\ref{tab1}.

\begin{table}[!htbp]
\centering
\caption{Mean values of different criteria for segmentation task.}\label{tab1}
\begin{tabular}{ccccccccccccccccc}
\hline
\specialrule{0em}{1.5pt}{1.5pt}
\multirow{2}{*}{Dataset} &\ & \multicolumn{3}{c}{Dice}&\ 
						 & \multicolumn{3}{c}{Sensitivity}&\ 
						 & \multicolumn{3}{c}{Specificity}&\ 
						 & \multicolumn{3}{c}{Hausdorff95}\\
						 &\ &ET&TC&WT&\ &ET&TC&WT&\ &ET&TC&WT&\ &ET&TC&WT\\\hline
						 \specialrule{0em}{1.5pt}{1.5pt}
Training &\ &0.830&0.888&0.916&\ &0.856&0.895&0.909&\ &0.998&0.997&0.996&\ &3.073&3.667&4.009\\
\specialrule{0em}{1.5pt}{1.5pt}
Validation &\ &0.737&0.807&0.894&\ &0.766&0.826&0.897&\ &0.998&0.996&0.995&\ &5.994&7.357&5.677\\
\specialrule{0em}{1.5pt}{1.5pt}
Final Test &\ &0.778&0.798&0.852&\ &-&-&-&\ &-&-&-&\ &3.543&6.219&6.547\\
\hline
\end{tabular}
\end{table}

For the OS days prediction task, only patients whose 'ResectionStatus' is 'GTR' are taken into consideration, which results in a 101 subjects training set, a 29 subjects validation set and a 107 subjects final test set. In this model, Adam optimizer with initial learning rate as 1e-4 has been deployed, it updates the learning rate the same way as in task one. For the training process, five-fold cross validation has been employed, which boils down to a configuration with batch size to be 5 and epochs to be more than 500.

Table~\ref{tab2} presents the scores of our predicted survival days. The accuracy is calculated based on a three categories classification, and they define the survival days less than 300 as short-survival, from 300 to 450 as mid-survivor, and more than 450 as long-survival.

\begin{table}[!htbp]
\centering
\caption{Mean values of different criteria for OS days prediction task.}\label{tab2}
\begin{tabular}{ccccccc}
\hline
\specialrule{0em}{1.5pt}{1.5pt}
Dataset&Accuracy&MSE&medianSE&stdSE&SpearmanR\\
\hline
\specialrule{0em}{1.5pt}{1.5pt}
Training& 0.515& 8.73e4&	1.96e4&	1.83e5&	0.472\\
Validation& 0.448&	1.0e5&	4.93e4&	1.35e5&	0.25\\
Final Test& 0.551&	4.10e5&	4.93e4&	1.23e6&	0.323\\

\hline
\end{tabular}
\end{table}

\section{Conclusion}
In this work, we introduced a brain-wise normalization and two patching strategies for the training of 3D U-Net for tumor segmentation task. At the same time, we brought about a network taking use of features extracted from predicted tumor labels to anticipate the overall survival days of patients who have undergone gross total resection surgery.  
Currently on single GPU platform, only one $4\times128\times128\times128$ image could be fed as input each time during training, which probably restricts the capacity of the model. In future works, with more powerful hardwares we would go on with the training and upgrading of this network.

\end{document}